\documentclass[]{aa}  
\usepackage{graphicx}
\usepackage{txfonts}
\usepackage{lipsum}
\usepackage{lscape}             
\usepackage{placeins}         
\usepackage[]{hyperref}

\begin{document}

\title{The Effect of Expansion and Instabilities in the Thermodynamic Regulation of the Young Solar Wind Plasma}
\author{M. Coello-Guzmán\inst{1,2}\corrauth{matilde.coello@usach.cl}       
    \and V. A. Pinto\inst{1,2}\email{victor.pinto@usach.cl}
    \and R. E. Navarro\inst{3}\email{roberto.navarro@udec.cl}
    \and P. S. Moya\inst{3}\email{pablo.moya@uchile.cl}
        }
   \institute{Departamento de Física, Universidad de Santiago de Chile, Santiago 9170124, Chile
   \and Center for Interdisciplinary Research in Astrophysics and Space Sciences (CIRAS), Universidad de Santiago de Chile, Santiago  9170124, Chile
   \and Departamento de F\'isica, Facultad de Ciencias F\'isicas y Matem\'aticas, Universidad de Concepci\'on, Concepci\'on 4070386, Chile
   \and {Departamento de Física, Facultad de Ciencias, Universidad de Chile, Santiago 7800003, Chile}}

   \date{Received April 01, 2026}
\titlerunning{Effect of expansion and instabilities in the young solar wind}
 
  \abstract
   {The solar wind ion temperature anisotropy with respect to the background magnetic field, $T_{\perp}/T_{\parallel}$, is constrained by $\beta_{\parallel}$-driven thermal microinstabilities. Above 0.3 Astronomical Units (au), the main active instabilities are oblique: mirror and firehose. In the same regions, the non-adiabatic expansion of the solar wind manifests in the empirical scaling $T_{\perp}/T_{\parallel}\sim\beta_{\parallel}^{-0.55}$ as radial distance grows.}
    {We extend these characterizations for the region under 0.13 au. We study the active microinstabilities constraining the margins of ion measurement distributions in the $(\beta_{\parallel}$, $T_{\perp}/T_{\parallel})$ parameter space, and how expansion effects drive the observable scaling between $T_{\perp}/T_{\parallel}$ and $\beta_{\parallel}$.} 
   {To achieve this, we organize Parker Solar Probe particle measurements in the $(\beta_{\parallel},\,T_{\perp}/T_{\parallel})$ parameter space, aggregating encounters that have reached perihelia under 0.1 au.}
   {At the surveyed radial distances, there is a clear dominance of parallel-propagating instabilities: electromagnetic ion-cyclotron and firehose. Furthermore, we show that the temperature anisotropy radially evolves following approximately the same semi-empirical anti-correlation $T_\perp/T_\parallel\sim \beta_\parallel^{-0.55}$ observed at greater distances.}
   {Under 0.1 au, the majority of $\beta_{\parallel}<1$ produces parallel-propagating instabilities.  As non-adiabatic expansion drives $\beta_{\parallel}$ values above unity, oblique instabilities are favored instead, in accordance with predictions from quasilinear kinetic theory. }

   \keywords{Solar wind --
            space plasmas --
            plasma physics -- 
            plasma astrophysics
               }
   \maketitle
\nolinenumbers

\section{Introduction}\label{sec:intro} 
The primary physical processes that control proton temperature anisotropies in the inner heliosphere remain an open question. The expansion of the solar wind and its marginal stabilization through kinetic instabilities are known fundamental drivers of temperature anisotropy. However, the coupling mechanism between them is not well understood~\citep{Matteini2011,Yoon2016,Yoon2019}. The Chew-Goldberger-Low (CGL) theory~\citep{CGL1956} predicts that a collisionless and adiabatically expanding plasma develops anisotropic temperatures that scale with the plasma beta as $T_{\perp}/T_{\parallel} \sim  \beta_{\parallel}^{-b}$ with $b=1$. Recent theoretical work incorporating radial expansion and Parker-spiral calculations refines the CGL picture, but is consistent with a steeper anti-correlation $b\geq1$~\citep{Seough2023,EcheverraVeas2024}. However, spacecraft measurements reveal systematic departures from CGL theory, with $b\simeq0.45-0.55$ for the fast solar wind between 0.29 and 0.98 astronomical units (au)~\citep{Marsch2004,Matteini2007}, suggesting that enhanced perpendicular heating or other non-adiabatic mechanisms are active in the explored regions. Early results from the Parker Solar Probe (PSP) mission indicate that fast winds below 0.24 au experience even stronger perpendicular heating~\citep{Huang2020,Huang2023}, with negligible net parallel heating below 0.326 au~\citep{Mozer2023}.

At 1 au, a significant fraction of solar wind measurements cluster near temperature isotropy and near-equipartition of thermal and magnetic pressures, suggesting the presence of additional thermalization processes~\citep{Hellinger2006,Bale2009,Maruca2011}. Meanwhile, the $T_\perp/T_\parallel$ measurement distribution tails for relatively high $\beta_\parallel$ seem to be constrained by the mirror and oblique firehose kinetic instabilities~\citep{Kasper2002,Hellinger2006,Matteini2007,Marsch2012,Matteini2013}. This picture is supported by observed enhancements in gyroscale magnetic fluctuation power both along these instability thresholds and in the regions subtended between thresholds, where scattered particles are detected~\citep{Bale2009,Maruca2011,Navarro2014}. Recent PSP analyses and kinetic simulations further identify instability signatures in the inner heliosphere, consistent with instability constraints near 0.3 au~\citep{Huang2020,Klein2021,Liu2023,Pezzini2024}. Additionally, numerical studies show that expanding plasmas are driven towards the firehose instability but are deflected backward once instabilities develop~\citep{Seough2013,Kunz2014,Yoon2019,Hellinger2019,Seough2023}. 

In this paper, we address how solar wind expansion couples with kinetic microinstabilities to regulate proton thermodynamics in the accelerating solar wind. To do this, we analyze Parker Solar Probe (PSP) particle and magnetic field measurements within $30$ solar radii ($R_\odot$) across Encounters 4 to 24, applying quality filters to remove partial field-of-view artifacts. By mapping these observations in the $(\beta_\parallel,\, T_\perp/T_\parallel)$ parameter space across radial distances and comparing them to linear kinetic growth-rate thresholds, we characterize the dominant microinstabilities constraining the young plasma and assess whether non-adiabatic expansion drives the empirical $T_\perp/T_\parallel \sim \beta_\parallel^{-b}$ scaling near the Sun.

\section{Data selection}

We employ measurements from the Solar Wind Electrons Alphas and Protons Investigation (SWEAP) instrument suite~\citep{Kasper2015}, particularly the Solar Probe Analyzers for ions (SPAN-i) electrometers. We use publicly available measurements of proton temperature tensors and proton density, and the local background magnetic field for encounters that reached perihelia under $30\,R_{\odot}$ (encounters 4 to 24, from January 2020 to June 2025). The bulk of the ion population sometimes flies past the SPAN-i detectors or is obstructed by the heat shield~\citep{Livi2022}; hence, the instruments may fail to capture the full distribution, thereby affecting density values, while temperature anisotropy behaves more robustly. The location of the detectors was designed to leverage the high tangential speed of the spacecraft near perihelia~\citep{Mostafavi2024}; thus, incomplete measurements are sparser under $30\,R_{\odot}$ but still prevalent enough to warrant filtering out data affected by partial field-of-view coverage~\citep{Ofman2022,Short2024,Yogesh2026}. Since the azimuth ($\phi$) angle is the most affected by heat-shield blocking in comparison to the polar ($\theta$) angle~\citep{Yogesh2026}, our dataset consists of measurements taken under $30\,R_{\odot}$, and where the coverage of at least 6 out of the 8 energy channels in the $\phi$ direction was ensured. With this procedure, $554\,112$ of the initial $2\,659\,276$ observations ($20.8\%$) were eliminated. Removing intervals with partial moments improves the statistical significance of our sample; this is evidenced in the noticeable organization exhibited by the aggregated measurements in Figures~\ref{fig:BA_all_data}, ~\ref{fig:ba_by_distance} and~\ref{fig:marsch_relation}, further discussed below.

\begin{figure}[ht!]
    \centering
\includegraphics[width=0.98\linewidth]{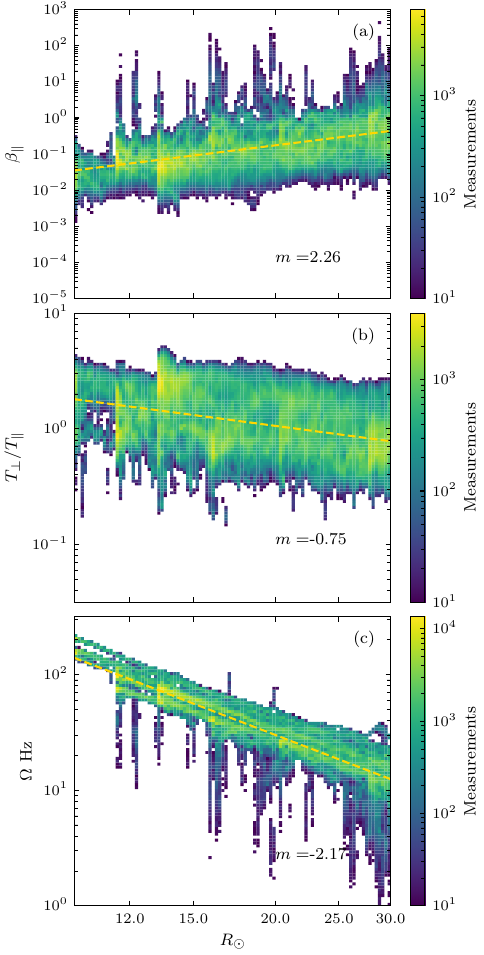}
    \caption{Bi-dimensional log-log histograms from PSP measurements across encounters 4-24, aggregated by distance. In each panel, the colorbar represents the net number of measurements counted in equal-sized bins for (a) the proton $\beta_\parallel$, (b) the proton temperature anisotropy $T_\perp/T_\parallel$, and (c) the proton gyrofrequency in Hertz. At the bottom right of each panel, $m$ is the slope of a linear fit, plotted in a dashed line.}
    \label{fig:figure1}
\end{figure}

\section{Radial trends} \label{sec:radtrends}

Figure~\ref{fig:figure1}(a) shows the radial profile of the proton $\beta_{\parallel}=8\pi n T_\parallel/B^2$, where $n$ is the proton density and $B$ the background magnetic field. Here, as the trend line suggests, $\beta_{\parallel}$ increases slightly as distance grows from $10$ to $30\,R_{\odot}$, indicating that magnetic confinement loses ground to kinetic agitation effects as the plasma expands, but remains the dominant driver as most measurements exhibit $\beta_{\parallel}<1$.

Figure~\ref{fig:figure1}(b) shows the evolution of temperature anisotropy. This change is both slower and the opposite of the trend portrayed by $\beta_{\parallel}$ during expansion. Most individual measurements show $T_\perp/T_\parallel>1$, implying that young solar wind protons are preferentially heated in the perpendicular direction. Finally, figure~\ref{fig:figure1}(c) shows the proton gyrofrequency $\Omega$ in Hz. Under a CGL approximation, we would expect $T_\perp/T_\parallel\sim\Omega$ and $\beta_\parallel\sim1/\Omega$ scalings as the solar wind expands~\citep{CGL1956}. However, the expected CGL scalings are not met by comparing the radial trends in Figs.~\ref{fig:figure1}(a)--(c), which is consistent with non-adiabatic expansion as previously observed for the solar wind beyond the acceleration region~\citep{Marsch2004,Matteini2007}. 

\begin{figure}[t!]
    \centering
    \includegraphics[width=0.98\linewidth]{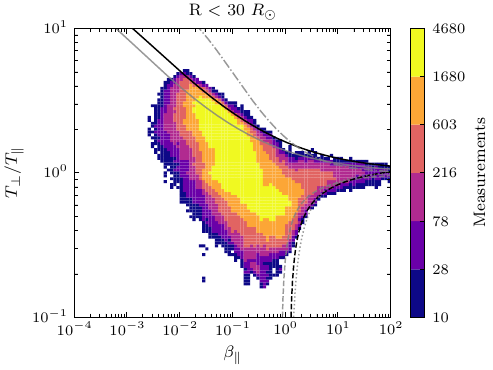}
    \caption{Color scale plot of the number of measurements of $(\beta_{\parallel},T_{\perp}/T_{\parallel})$, with bins in log-log scale. Selected measurements from encounters 1-24 are used. Instability thresholds for the EMIC, mirror, parallel firehose, and oblique firehose instabilities are shown with solid, dash-dotted, dashed, and dotted lines, respectively, with contours $\gamma_{\text{max}}/\Omega=10^{-3}$ in gray, and $10^{-2}$ in black.}
    \label{fig:BA_all_data}
\end{figure}

\begin{figure*}[ht!]
    \centering
    \includegraphics[width=0.98\linewidth]{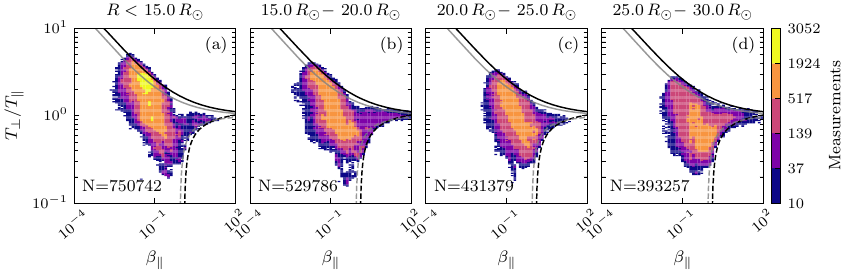}
    \caption{Measurements organized in the $(\beta_{\parallel}, T_{\perp}/T_{\parallel})$ plane filtered by radial distance. $N$ is the total number of measurements in each panel. The shared colorbar represents the counts per bin on a log-scale. Only the EMIC and parallel firehose instability thresholds are shown for $\gamma_{\text{max}}/\Omega=10^{-3}$ (gray lines) and $10^{-2}$ (black lines).}
    \label{fig:ba_by_distance}
\end{figure*}

Figure~\ref{fig:BA_all_data} shows the same observations as in Figure~\ref{fig:figure1}, but organized in the $(\beta_{\parallel},\, T_{\perp}/T_{\parallel})$ parameter space. We also plot with gray lines the mirror and electromagnetic ion-cyclotron (EMIC) instability thresholds for $T_{\perp}/T_{\parallel}>1$, and the parallel and oblique firehose instability thresholds for $T_{\perp}/T_{\parallel}<1$~\citep{Gary1993,Hellinger2006,Yoon2017}, all of them corresponding with the maximum growth rate $\gamma_{\text{max}}/\Omega=10^{-3}$ which is usually associated with marginal regulation at 1 au~\citep{Hellinger2006, Bale2009, Maruca2011}. Figure~\ref{fig:BA_all_data} shows that the observed $T_{\perp}/T_{\parallel}>1$ upper bounds are better aligned, although not fully constrained, with the EMIC instability. This contrasts with what is observed at 1 au, where the non-propagating mirror and oblique firehose modes are dominant~\citep{Kasper2002,Bale2009,Maruca2011}. Thus, Fig.~\ref{fig:BA_all_data} suggests that mostly parallel propagating instabilities may be dominant in the acceleration region.

Under $30\,R_{\odot}$ the background magnetic field appears mostly radial~\citep{Zhao2025,Bian2024}, a feature that should privilege parallel propagating electromagnetic waves~\citep{Jian2009,Wicks2016,Zhao2021}. Indeed, early observations~\citep{Jian2009,Verniero2020} and numerical kinetic modeling~\citep{Ofman2025} suggest that the inner-heliospheric solar wind can host quasi-parallel, circularly polarized ion-scale electromagnetic waves, including Alfvén/ion-cyclotron modes driven by free energy in non-Maxwellian proton distributions. 

Since a fraction of the measurements lie above the commonly overplotted $\gamma_{\text{max}}/\Omega=10^{-3}$ level used to represent marginal stability at 1~au (seen in gray in Fig.~\ref{fig:BA_all_data}), we numerically compute the linear EMIC growth-rate from the proton dispersion relation~\citep{Navarro2014,Swanson1989-cb} for different values of $\beta_{\parallel}$ and $T_{\perp}/T_{\parallel}$, and identify the mode exhibiting the largest growth-rate. We then define an empirical stability boundary as the contour where $\gamma_{\text{max}}/\Omega=10^{-2}$, which is plotted in Fig.~\ref{fig:BA_all_data} with black lines. As shown, this contour provides a markedly improved envelope for the observed distribution compared to the $\gamma_{\text{max}}/\Omega=10^{-3}$ level. A non-linear least-squares fit with the generalized power-law $T_\perp/T_\parallel=A_0+a\,(\beta_\parallel-\beta_0)^{-c}$, where $A_0$, $a$, $\beta_0$, and $c$ are free fitting parameters, results in $A_0=1.02$, $a=0.62$, $\beta_0 =-0.0002$, and $c=0.41$ for the EMIC instability, and $A_0=1.15$, $a=-0.74$, $\beta_0 = 0.91$, and $c=0.37$ for the parallel firehose instability. These fitting parameters are close to those reported by~\citet{Verscharen2019} (their Table 3), although we added a fourth parameter, $A_0$.

The specific choice of a finite growth-rate level remains empirical. In the inner heliosphere, the solar wind is an open, accelerating, and expanding system that is generally far from local thermodynamic equilibrium, so there is no a priori requirement that the data must be bounded by thresholds derived under near-equilibrium, homogeneous, and typically idealized assumptions (e.g., bi-Maxwellian protons). Nevertheless, the strong measurement organization under the EMIC instability threshold exhibited in Figure~\ref{fig:BA_all_data} suggests that unstable parallel-propagating ion-cyclotron waves are dynamically relevant for the very core of the distribution. At 1~au, EMIC and mirror contours corresponding to a $\gamma_{\text{max}}/\Omega=10^{-2}$ growth rate could be relevant to the margin of the distribution tails, even considering the considerable scattering beyond the instability thresholds~\citep{Maruca2011, Verscharen2019}. In contrast, Figure~\ref{fig:BA_all_data} shows that under $30\, R_{\odot}$ only $\sim0.4\%$ of the observations surpass the EMIC threshold, further supporting the importance of unstable parallel-propagating waves and this particular timescale. One self-consistent interpretation for this is that only sufficiently fast-growing modes (greater than $\gamma_{\text{max}}/\Omega=10^{-3}$) can act effectively within the finite time and spatial extent over which unstable conditions persist, thereby setting an effective marginality level that differs from the conventional 1~au choice.

\section{The role of expansion} \label{sec:expansion}

To characterize the interplay between instability regulation and plasma expansion, we show in Fig.~\ref{fig:ba_by_distance} the $(\beta_\parallel,\, T_\perp/T_\parallel)$ parameter space distribution organized by radial distance from the Sun. Here, we only show the parallel EMIC and firehose instabilities, and we see that these curves constrain the data at all distances below $30\, R_\odot$. Considering that Fig.~\ref{fig:ba_by_distance}(a) has the most net measurements ($N$ at the bottom left of each panel), the sparsely populated region close to the firehose-type instability region signals that this instability certainly is active, although not playing the predominant role in regulating the plasma. Since the bulk of the particles concentrates in the $T_{\perp}/T_{\parallel}>1$, low-$\beta_{\parallel}$ region, they are more likely to excite the EMIC mode. Indeed, the very centroid of the distribution (bins containing over $517$ counts) exhibits a well-defined instability constraint on its $T_\perp/T_\parallel$ upper bound near $\beta_\parallel\sim10^{-1}$. Also, for the distance interval under $15\,R_{\odot}$, the distribution displays a sharper maximum (bins above 1924 counts) not present at the following distances, probably due to the isotropizing effects of expansion; this comparative feature is best illustrated by the shared colorbar. From Figs.~\ref{fig:ba_by_distance}(b)-(c)-(d), the expansion pushes towards the region with higher $\beta_{\parallel}$ and dominant parallel temperature eventually drives firehose modes, a behavior consistent with quasilinear theoretical predictions for an expanding collisionally regulated turbulent plasma~\citep{Yoon2019,Hellinger2019}. In Figs.~\ref{fig:ba_by_distance}(c) and (d), the bulk of the distribution concentrates around temperature isotropy and bulges near the firehose-instability region, implying that mechanisms counter-balancing the expansion, likely marginal deflection by the firehose instability~\citep{Yoon2017b,Yoon2019}, are active in the young solar wind.

Quasilinear simulations investigating the competition between EMIC and mirror instabilities~\citep{Yoon2012} have shown that the resulting proton dynamics is mostly driven by $\beta_{\parallel}$. For $\beta_{\parallel}\geq1$, both mirror and EMIC modes equally develop in the earlier stages, but the mirror mode quickly becomes dominant as $\beta_\parallel$ grows. In contrast, for $\beta_{\parallel}<1$, EMIC is the initial dominant mode. Since the PSP observations presented here show that $\beta_\parallel$ is low near the Sun, it is expected that EMIC resonances markedly dominate over mirror modes, explaining the measurements of the inner heliosphere presented in Figures~\ref{fig:BA_all_data} and \ref{fig:ba_by_distance}.

Beyond the inner heliosphere, the empirical anisotropy bounds transition from being most consistent with EMIC and parallel firehose constraints to being more consistent with oblique mirror and oblique firehose constraints~\citep{Hellinger2006,Huang2020,Huang2025}, suggesting that the regulating microinstability changes as the solar wind plasma evolves during expansion. Although microinstabilities provide a strong negative feedback that tends to keep the plasma near a marginally stable state~\citep{Yoon2012}, other processes, such as stochastic heating, can push the plasma towards unstable conditions~\citep{Bowen2025}. Indeed, a sufficient turbulent magnetic power can drive transverse heating and increase anisotropy~\citep{Moya2021}, pushing marginally stable parcels towards states consistent with the mirror instability and broadening the anisotropy distribution. In this view, the observed bounds may shift due to the interplay between turbulent expansion-driven dynamics and instability-mediated relaxation. 

\begin{figure}[ht!]
    \centering
    \includegraphics[width=0.98\linewidth]{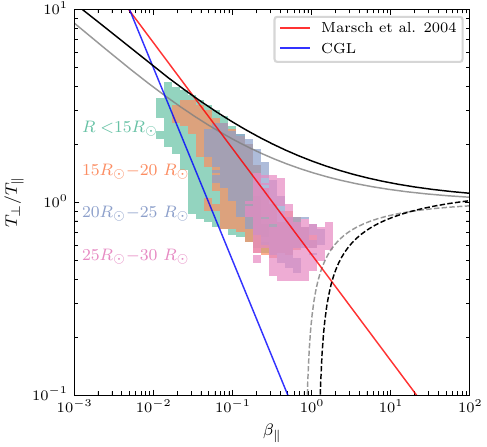}
    \caption{Centroids for each panel in Figure~\ref{fig:ba_by_distance}, corresponding to bins with over 517 measurements. The red line is~\citet{Marsch2004} relation $T_{\perp}/T_{\parallel}\sim\beta_{\parallel}^{-0.55}$. The blue line is the CGL prediction $T_{\perp}/T_{\parallel}\sim\beta_{\parallel}^{-1}$}
    \label{fig:marsch_relation}
 \end{figure}

Finally, Figure~\ref{fig:marsch_relation} presents the centroids of the distributions shown in Figure~\ref{fig:ba_by_distance} along the CGL-derived prediction in blue, and the semi-empirical relation  $T_\perp/T_\parallel\sim \beta^{-b}$ with $b=0.55$ proposed by~\cite{Marsch2004} in red. This trend was superimposed using the bin with the highest count in Figure~\ref{fig:ba_by_distance}(a) as the starting point. The semi-empirical trend is remarkably followed. Not only do the margins of the centroids in the studied distances follow the EMIC instability thresholds, but the scaling proposed by~\cite{Marsch2004} is consistently observed. A comparison with studies at distances beyond $60\, R_{\odot}$~\citep{Matteini2007} highlights a seemingly universal scaling law from 0.1 to 1 au. Figures~\ref{fig:ba_by_distance} and~\ref{fig:marsch_relation} demonstrate that the bulk of the solar wind protons within $15-30\, R_{\odot}$ undergo a non-adiabatic expansion that diverts from CGL predictions, as the borders of the distribution are subject to parallel-propagating thermal microinstability regulating processes.

\section{Discussion and conclusions} \label{sec:summary}

In the present work, we show that temperature anisotropies observed by PSP inside $30\, R_{\odot}$ are not constrained by the mirror and oblique firehose thresholds that bound the solar wind at 1 au. Instead, PSP measurements show that parallel instabilities actively regulate the proton temperature anisotropy closer to the Sun, in accordance with predictions of quasilinear theory~\citep{Yoon2012,Yoon2019}. In light of earlier results for further distances~\citep{Hellinger2006,Matteini2007,Bale2009, Maruca2011}, our findings differentiate two stages for the marginal stability evolution in the $(\beta_{\parallel},\, T_{\perp}/T_{\parallel})$ parameter space: In the highly energetic, anisotropic young solar wind, more unstable parallel instabilities develop, gradually saturate, and are then overtaken as the plasma expansion drives $\beta_{\parallel}$ growth; and beyond a certain distance, once enough particles exhibit $\beta_{\parallel}\geq 1$,  non-propagating oblique instabilities (mirror and oblique firehose) become dominant. The transition occurs beyond $30\,R_{\odot}$. We also show that the temperature anisotropies undergo the same non-adiabatic expansion scaling $T_\perp/T_\parallel\sim\beta_\parallel^{-0.55}$ observed beyond the acceleration regions up to 1 au, even though asymptotic velocities are not yet reached under $30\, R_{\odot}$.

Theoretical findings suggest that the description of instability thresholds in the acceleration regions could be further improved by considering drivers such as electron-proton collisionality, compressive large-wavelength fluctuations, or by varying the assumed particle velocity distribution function~\citep{Yoon2017b,Verscharen2016, Yoon2024,Isenberg2013}. Recent PSP observations also support the stabilizing role of alpha particles and consider the free energy available in non-thermal features different than temperature anisotropy, such as beams related to secondary ion populations~\citep{Xiang2023,McManus2024,Niranjana2026}. Therefore, a multi-species model should be necessary to further characterize the role of non-thermal features and kinetic instabilities on the thermodynamic regulation of the solar wind. The relative importance of these features for the regulation of measurable quantities in the young solar wind plasma can only be gauged through careful future analysis and interpretation of observations from spacecraft such as PSP. Our work is concerned with the role of $\beta_{\parallel}$ as the main driver that determines which type of marginal instability regulation is more prominent as the solar wind undergoes non-adiabatic expansion, parallel near the Sun, and oblique at 1 au.

\begin{acknowledgements}
We are grateful for the support of DICYT Project 042431PA\_Postdoc (MCG, VAP) of Universidad de Santiago de Chile, and ANID Chile through the FONDECYT grants No.  3260929 (MCG), No. 11251905 (VAP), No. 1240281 (PSM), and No. 1240697 (REN). We acknowledge the support of the computing infrastructure at USACH through the Fondequip EQM230160 grant and the CIRAS-AI FIUF137139-USACH project. This research was supported by the International Space Science Institute (ISSI) in Bern, through the ISSI International Team project 24-612: Excitation and Dissipation of Kinetic-Scale Fluctuations in Space Plasmas. We acknowledge the NASA Parker Solar Probe Mission and the SWEAP team led by J. Kasper for the use of data.
\end{acknowledgements}

\bibliographystyle{aa} 
\bibliography{refs_AA}

\end{document}